\documentclass[twocolumn,showpacs,preprintnumbers,amsmath,amssymb]{revtex4}
\usepackage{graphics}
\usepackage[dvips]{graphicx}
\usepackage{color}
 
\def\setR{\mathbb{R}}

\def\ie {i.e.} 

\DeclareMathAlphabet{\mcal}{OT1}{rsfs}{m}{sl}

\newcommand{\sss}[1]{\scriptscriptstyle #1} 

\begin{document}

\title{Revisiting the conformal invariance of the scalar field:\\
from Minkowski  space to de Sitter space}

\author{
E. Huguet$^1$, J. Queva$^1$, J. Renaud$^2$
}
\affiliation{$1$ - Universit\'e Paris 7-Denis-Diderot, APC-Astroparticule et Cosmologie (UMR 7164), 
Batiment Condorcet, 10 rue Alice Domon et L\'eonie Duquet, F-75205 Paris Cedex 13, France.  \\
$2$ - Universit\'e Paris-Est, APC-Astroparticule et Cosmologie (UMR 7164), 
Batiment Condorcet, 10 rue Alice Domon et L\'eonie Duquet, F-75205 Paris Cedex 13, France.
} 
\email{huguet@apc.univ-paris7.fr, queva@apc.univ-paris7.fr, jacques.renaud@univ-mlv.fr}

\date{\today}

\begin{abstract}
In this article,
we clarify the link between the conformal (\ie~Weyl) correspondence from 
the Minkowski space to the de Sitter space and the conformal (\ie~SO(2,$d$)) invariance 
of the conformal scalar field  
on both spaces. We exhibit the realization on de Sitter space of the 
massless scalar 
representation of SO$(2,d)$. It is obtained from the corresponding representation in 
Minkowski space through an intertwining operator inherited from the Weyl relation between the two spaces. 
The de Sitter representation is written in a form which allows one to take the point of view of a Minkowskian observer who sees the effect of curvature through additional terms.

\end{abstract}

\maketitle

\section{Introduction}
In this paper we are interested in the conformally coupled massless scalar field (hereafter conformal 
scalar field, for simplicity) in both $d$-dimensional 
Minkowski and de Sitter spaces for $d>2$. It is a known fact that each of these two fields, Minkowskian and 
de Sitterian, are  invariant under the group SO$(2,d)$. It is also well-known that these two spaces are related through a Weyl rescaling. These properties are rarely used at the same time (see however \cite{KB}) and are both termed 
``conformal invariance" in the literature. In this paper, we show how they hang together. In short, the Weyl rescaling induces a map $\widehat{\Xi}_{\sss H}$ between the two Hilbert spaces of solutions of the corresponding conformal equations. This map intertwines the two representations carried by these spaces.

The operator $\widehat{\Xi}_{\sss H }$ was introduced, in substance,  in a previous article \cite{pconf1}. In that work  
we realized Minkowski, de Sitter, and anti-de Sitter  
spaces on the same underlying set. Since the anti-de Sitter space is not globally hyperbolic we restrict our present investigation  to the $d$-dimensional Minkowski and de Sitter spaces.
We used the conformal (Weyl) relation between them to 
deform Minkowskian objects into de Sitterian objects. In particular, it turns out that the 
operator $\widehat{\Xi}_{\sss H }$ is unitary between the Hilbert spaces of solutions of 
the conformal scalar field equations on both de Sitter and Minkowski spaces. 
As remarked, since the respective isometry groups of Minkowski and de Sitter spaces are different, one cannot
speak about the covariance of the operator $\widehat{\Xi}_{\sss H }$ with respect to these groups. Nevertheless, since these two spaces admit the same conformal group SO$(2,d)$, in which both Poincar\'e and 
de Sitter groups are included, the operator $\widehat{\Xi}_{\sss H }$ makes explicit 
the link between the two actions of SO$(2,d)$  
on Minkowski and de Sitter spaces. More precisely, this operator intertwines these representations.

Even more, this operator allows an explicit realization of the generators of the so$(2,d)$ algebra for the conformal field on the de Sitter space. This realization is written in a Minkowskian form: in some sense we deal with exact de Sitterian objects (without any approximation) in Minkowski space.

We organize our paper as follows. In Sec.~II we comment about the two kinds of conformal invariance.
The action of SO$(2,d)$ on Minkowski space is reminded in Sec.~III.
In Sec.~IV the Weyl rescaling between Minkowski and de Sitter space  is recovered
by using the
results of \cite{pconf1}. This allows to define the operator $\widehat{\Xi}_{\sss H }$. This operator is used
in Sec.~V to move the SO$(2,d)$ representation. Our conventions and some details  concerning the so$(2,d)$ algebra are given in the appendix.

\section{About conformal invariance}

Two notions of conformal invariance are used in this work; let us first make clear the
distinction between them.

The first one is connected with the conformal group SO$(2,d)$ 
\cite{barutraczka, MS}. 
Let us consider an equation for some field $F$ and write it symbolically as $O F = 0$, 
where $O $ denotes some linear operator. Such an equation is said 
 to be invariant under SO$(2,d)$, and we will keep this terminology in the sequel, 
if one can realize the generators 
$X_{\alpha\beta}$ of the Lie algebra so$(2,d)$ in such 
a manner that $[O, X_{\alpha\beta}] = \zeta O$, $\zeta$ being some function. The space of 
solutions of the equation $O F = 0$ is, in this case,  invariant under the corresponding action of the group SO$(2,d)$ and thus carries a representation of this group.

The second notion of conformal invariance is related to 
the so-called Weyl rescaling \cite{wald, Fulton:1962bu}. 
This transformation consists in a change from the metric $g$ defined on some
manifold $M$ to a new one $\overline{g} := \omega^2 g$, $\omega$ being a real function over $M$. 
Let us consider again the equation $O F = 0$; note that $O$ generally depends on $g$. This equation is said
to be conformally invariant, if there exists a number $s \in \setR$ (the conformal weight of the field) 
such as $F$ is a solution with $g$ {\it iff} $\overline{F} := \omega^s F$ is a solution with $\overline{g}$.
In all cases considered here, one has in fact  
$\overline{O}~\overline{F} = \omega^{s'} O F$, for some $s' \in \setR$. In the sequel, this kind
of conformal transformation and invariance will be referred to as ``Weyl transformation" and ``Weyl invariance".

It is well known that Minkowski and de Sitter spaces have the same conformal group, namely
SO$(2,d)$, whose Poincar\'e group SO$_0(1,d-1) \ltimes \setR^d$, and de Sitter group  SO$_0(1,d)$, are the respective subgroups of isometry.
In addition, Minkowski and de Sitter spaces are 
related by a Weyl transformation. 

One can 
ask how the SO$(2,d)$ invariance of the conformal field equation is ``transported" under the  Weyl transformation? That question is explicitly answered in this work. 

\section{The action of SO$(2,d)$ on the Minkowskian scalar field}

Now, we want to comment about the scalar representation of SO$(2,d)$ on Minkowski space which realizes the  
SO$(2,d)$ invariance. 

The conformal group of Minkowski space, whose linear representation is SO$(2,d)$, is given by its isometries
extended by the dilations and the special conformal transformations.
These transformations act, respectively, on the Minkowski 
coordinates $x^\mu$ as 
\begin{align}
&y^\mu_{\sss D} = \lambda x^\mu,\label{Dcoor}\\
&y^\mu_{\sss SCT} = \frac{x^\mu + b^\mu x^2}{1 + 2 b\!\cdot\! x + b^2 x^2},\label{SCTcoor}
\end{align} 
$\lambda$ and $b$ being parameters of the transformations. The well-known \cite{barutraczka} representation of the so$(2,d)$ algebra follows
\begin{align*}
&M_{\mu\nu}=x_{\mu}\partial_{\nu}-x_{\nu}\partial_{\mu},\\
&P_\mu=\partial_{\mu},\\
&D=x^{\mu}\partial_{\mu},\\
&K_\mu=2x_\mu x^{\nu}\partial_{\nu}-x^2\partial_{\mu}.
\end{align*}

One could think, and it can be found in some literature, that for any generator $X$ in this representation one has $[\Box,X]=\zeta\, \Box$. {\em This is not the case}.
In order to understand this fact, let us consider the usual action $S[\phi]$ for the conformal scalar field 
in Minkowski space (that is the free massless scalar field).
It is of course invariant under Poincar\'e transformations, but not under the above (\ref{Dcoor},\ref{SCTcoor}) transformations. 
In order to restore the action invariance, the field must be changed by an internal transformation a 
so-called scaling \cite{felsager}: the change of variables (\ref{Dcoor},\ref{SCTcoor}) must be performed
together with the change of function
$\phi' = \omega^s  \phi$ 
with
$
s = -(d - 2)/2$, 
$\omega_{\sss D}= \lambda$, and
$\omega_{\sss SCT}= (1 - 2 b \cdot  x + b^2 x^2)$ respectively. 
As a consequence, the action on
 the conformal scalar field is given by
\begin{equation}\label{action}
(T_g\phi)(x)=\omega_g^{-\frac{d-2}{2}}(x)\phi(g^{-1}x).
\end{equation}
Note that, the scaling term in (\ref{action}) is essential to make the above representation unitary 
with respect to the Klein-Gordon scalar product. 
When $g$ is an isometry, the scaling term $\omega_g$ is equal to 1.
 For $g$ 
a dilation or a special conformal transformation the equation (\ref{action}) specializes into
\begin{align*}
&T_{\sss D}\phi(x)
=\lambda^{-\frac{d-2}{2}}\phi\left(\frac{x}{\lambda}\right) ,\\
&T_{\sss SCT}\phi(x)=\\
&\hskip20pt\left(1 - 2 b\!\cdot\!x + b^2 x^2\right)^{-\frac{d-2}{2}}
\phi\left( \frac{x^\mu - b^\mu x^2}{1 -2 b\!\cdot\!x + b^2 x^2}\right).
\end{align*}

The corresponding representation of the Lie algebra is obtained by differentiation as usual:
\begin{align*}
&{M}^0_{\mu\nu}=x_{\mu}\partial_{\nu}-x_{\nu}\partial_{\mu}, \\
&{P}^0_\mu=\partial_{\mu},\\
&{D}^0= \frac{d-2}{2} + x\cdot\partial,\\
&{K}_\mu^0 = (d-2) x_\mu + (2x_\mu x^\nu - x^2 \delta_\mu^\nu)\partial_\nu,
\end{align*} 
where the exponent 0 reminds that we are dealing with the Minkowski case.
These generators can also be found from group theory arguments \cite{MS}.
They  verify 
\begin{align*}
&[\square,{P}_\mu^0]=[\square, {M}_{\mu\nu}^0]=0,\\
&[\square, {D}^0] = 2 \square, \\
&[\square,{K}_\mu^0] = 4 x_\mu \square.
\end{align*} 

Before closing this section we want to emphasize that the conformal scalar field is not a scalar field 
in the sense of the differential geometry.  Remind that, a complex map 
$\varphi$ defined on $\setR^d$ is a scalar field iff for any diffeomorphism $f$ defined on $\setR^d$ one
has 
$(f_*\varphi)(f(x)) = \varphi(x)$. In this respect, the conformal scalar field in 
Minkowski space is not really a scalar field. In fact, it is a scalar field only for Poincar\'e transformations.
The same remark extends to the conformal scalar field  in de Sitter space. 

\section{The cone up to the dilations and the Weyl rescaling}\label{papier1}

In this section we use the results of our previous work \cite{pconf1}, to which we refer the reader for further details, to define the $\widehat{\Xi}_{\sss H}$ map. This is done by first retrieving the Weyl relation 
between Minkowski and de Sitter spaces. 

Both Minkowski and de Sitter spaces can be obtained as the intersection of the null cone of $\setR^{d+2}$ and a moving hyperplane (see the appendix for the notations). Precisely, the manifold
\begin{equation*}
X_{\sss H} = \mathcal{C} \cap P_{\sss H},
\end{equation*}
where
\begin{equation*}
\mathcal{C} = \left\{y \in \setR^{d+2} :  (y^{d+1})^2 + (y^0)^2 - \boldsymbol{y}^2 - (y^d)^2 = 0\right\}, 
\end{equation*}
and
\begin{equation*}
P_{\sss H} = \left\{y \in \setR^{d+2} : (1 + H^2) y^{d+1} + (1 - H^2) y^{d} = 2 \right\}, 
\end{equation*}
can be shown to be the Minkowski or a de Sitter space for $H = 0$ and $H > 0$ respectively. In addition,
one can show that $X_{\sss H}$ can be seen as a subset of the cone up 
to the dilations $\mathcal{C}'=\mathcal{C}/\sim,$ where the relation $\sim$ is defined through $u\sim v$ 
iff
there exists $\lambda>0$ such that $u=\lambda v$. From this point of view $X_{\sss H}$ is defined through
\begin{equation}
X_{\sss H}= \left\{y \in \mathcal{C}' : (1 + H^2) y^{d+1} + (1 - H^2)y^d > 0\right\}. 
\end{equation}
Let us reemphasize that both Minkowski and de Sitter spaces are realized on the same underlying set. 

The line elements 
$ds_{\sss H}$ on $X_{\sss H}$ and $ds$ on $\mathcal{C}'$ are conformally related,~\ie, 
\begin{equation*}
ds_{\sss H}^2 = \Omega_{\sss H}^2 ds^2.
\end{equation*}
From this relation we deduce 
\begin{equation*}
ds_{\sss H}^2 = \Xi_{\sss H}^2 ds^2_0,
\end{equation*}
where
\begin{equation*}
\Xi_{\sss H} := \frac{\Omega_{\sss H}}{\Omega_0},
\end{equation*}
which is the Weyl relation between de Sitter and Minkowski spaces.

Owing to this 
Weyl transformation 
the solutions of both conformal scalar field equations on Minkowski and de Sitter spaces 
are obtained from the conformal scalar field equation on $\mathcal{C}'$ which is simpler. Moreover, 
let ${\cal H}_0$ and $\mathcal{H}_{\sss H}$ be the Hilbert spaces of the solutions, square integrable 
with respect to the Klein-Gordon scalar product, of the Minkowskian 
(respectively de Sitterian) conformal scalar field equation, the map 
\begin{equation*}
\begin{array}{lrcl}
\widehat{\Xi}_{\sss H} : &{\cal H}_0 &\to& \mathcal{H}_{\sss H}\\
&\phi&\mapsto& \widehat{\Xi}_{\sss H}(\phi) :=\Xi_{\sss H}^{-(\frac{d-2}{2})}\phi,
\end{array}
\end{equation*}
is unitary. Now, let us consider some
well defined linear operator $O^0$ over ${\cal H}_0$, the unitarity of 
$ \widehat{\Xi}_{\sss H}$
ensure that the
moved operator
\begin{equation}\label{defOH}
O^{\sss H} :=  \widehat{\Xi}_{\sss H} O^0  \widehat{\Xi}_{\sss H}^{-1},
\end{equation}
is well defined over ${\cal H}_{\sss H}$.

Since Minkowski and de Sitter spaces are written on the same underlying set one can use a common system of coordinates. A convenient one, for our purpose, is the usual Minkowski system $\{x^\mu\}$ in which 
\begin{equation*}
\Xi_{\sss H} = \frac{1}{1 - \frac{H^2}{4} x^2}.
\end{equation*}

\section{The SO$(2,d)$ covariance of the $ \widehat{\Xi}_{\sss H}$ map}

Having defined the $ \widehat{\Xi}_{\sss H}$ map, we show how it connects the SO$(2,d)$ invariance of 
the two conformal scalar fields. 

As recalled above, $ \widehat{\Xi}_{\sss H}$ is a unitary operator between the Hilbert spaces of solutions of the conformal scalar field equation on Minkowski and de Sitter spaces.
The group acting on de Sitter and Minkowski spaces are different; as a consequence, there is no meaning to speak about covariance of $\widehat{\Xi}_{\sss H}$ with respect to these groups. 
We thus focus on the conformal group SO$(2,d)$ which is the same 
for both spaces. 

Since these spaces
are realized on the cone up to the dilations on which SO$(2,d)$ has a natural action, the geometric action is the same for the two spaces at least locally.
As a consequence, the term $\phi(g^{-1}x)$ in (\ref{action}) is the same in both Minkowski and de Sitter spaces.
The scaling term $\omega_g(x)$ will, however, be different, depending  
on $H$ and denoted
$\omega^{\sss H}_g(x)$ in the following.
 One could compute this term by considering the action $S[\phi]$ of the field, but it can be more easily recovered 
by moving the representation from Minkowski space to de Sitter space using (\ref{defOH}).
A straightforward computation shows that the  de Sitter scaling term $\omega_g^{\sss H}(x)$ reads for any $g\in$ SO$(2,d)$
\begin{equation*}
\omega_g^{\sss H}(x)=\left(\Xi_{\sss H}(x)\right)^{-\frac{2}{d-2}}\omega_g(x)
\left(\Xi_{\sss H}(g^{-1}x)\right)^{\frac{2}{d-2}}.
\end{equation*}
The generators, in Minkowskian coordinates,  follow in the same way
\begin{align}\label{genetransportes1}
{M}_{\mu \nu}^{\sss H}& = {M}_{\mu \nu}^0, \\
{P}_\mu^{\sss H} &= {P}_\mu^0 +\frac{d-2}{2}(\partial_\mu \ln \Xi_{\sss H})\label{genetransportes2} \\ 
&=\partial_\mu+\frac{d-2}{4}H^2\frac{x_\mu}{1-\frac{H^2}{4}x^2},\label{genetransportes3}\\
{D}^{\sss H} &= {D}^0 + \frac{d-2}{2}(x^\mu\partial_\mu \ln \Xi_{\sss H}) \label{genetransportes4}\\
&= x\cdot\partial +\frac{d-2}{2} \left(\frac{1+\frac{H^2}{4}x^2}{1-\frac{H^2}{4}x^2}\right),\label{genetransportes5}\\
{K}_\mu^{\sss H} &= {K}_\mu^0 + \frac{d-2}{2}(2x_\mu x^\nu - x^2 \delta_\mu^\nu)\partial_\nu \ln \Xi_{\sss H}
\label{genetransportes6}\\
&=(2x_\mu x^\nu - x^2 \delta_\mu^\nu)\partial_\nu+(d-2)\frac{x_\mu}{1-\frac{H^2}{4}x^2}.
\label{genetransportes7}
\end{align}
 These generators satisfy by construction
the so$(2,d)$ algebra (\ref{so2debut}-\ref{so2dfin}).
As expected, the operators 
${P}_\mu^0$ 
are moved into non-isometric operators; this is confirmed by the presence of the non-derivative term. The de Sitter isometries are identified at the lie algebra level (\ref{yh}) and then 
realized on de Sitter space
through (\ref{genetransportes1}-\ref{genetransportes7}). They read
\begin{eqnarray*}
Y_\mu^{\sss H} &=& {P}_\mu^{\sss H} - \frac{H^2}{4} {K}_\mu^{\sss H}\\
&=& {P}_\mu^{0} - \frac{H^2}{4} ({K}_\mu^0- (d-2)x_\mu)\\
&=&\partial_\mu-\frac{H^2}{4}(2x_\mu x^\nu - x^2 \delta_\mu^\nu)\partial_\nu .
\end{eqnarray*}
This can be interpreted as the expression of the infinitesimal translation 
from the point of view of a Minkowskian observer living in a de Sitter space:
the Minkowskian translation plus a term due to the curvature.

Note that the combination $I_\mu^{\sss H}:= {K}_\mu^{\sss H} - x^2 {P}_\mu^{\sss H}$ is invariant under $\widehat{\Xi}_{\sss H}$, that is $[\widehat{\Xi}_{\sss H},I_\mu^{\sss H}] = 0$ for
all values of $H$. 

We finally comment about the expression of the conformal field equation in de Sitter space ($H>0$) which 
reads $O \phi^{\sss H} = 0$, with $O :=(\square_{\sss H} + d (d-2) H^2 /4 )$.
One can be tempted to identify the operator $O$
with the moved d'Alembertian ${\displaystyle O':=\widehat{\Xi}_{\sss H} \square \,\widehat{\Xi}_{\sss H}^{-1}}$. 
In fact a direct calculation shows that 
\begin{equation*}
 O = \Xi^{-\frac{2d}{d-2}}_{\sss H} O'.
\end{equation*}
Of course, both operators equal to zero on ${\cal H}_{\sss H}$. 

\section*{ACKNOWLEDGMENTS}
The authors thank G. Decerprit for his helpful contribution and M. Lachi\`eze-Rey for useful discussions.  

\appendix
\section{}
Here are the conventions about indices:
\begin{eqnarray*}
\alpha, \beta, \gamma, \delta, \ldots &=&d+1, 0, \ldots, d,\\
\mu,\nu,\rho,\sigma,\ldots &=&  0, \ldots, d-1,\\
i, j, k, l, \ldots &=& 1, \ldots, d-1.
\end{eqnarray*} 
The coefficients of the metric of $diag(1,1,-1,\ldots,-1)$ of $\setR^{d+2}$ are denoted $\eta_{\alpha \beta}$:
\begin{equation*}
\eta_{d+1\ d+1}=\eta_{00}=1=-\eta_{ii}=-\eta_{dd}.
\end{equation*}
In addition, when no confusion is possible we use the  superscript $0$ ($H$) to denote
a Minkowskian (de Sitterian) quantity. 

The generators of the algebra so$(2,d)$ are 
$X_{\alpha \beta}~=~y_\alpha \partial_\beta -  y_\beta \partial_\alpha$, they satisfy
\begin{equation*}
\left[X_{\alpha \beta}, X_{\gamma \delta}\right] =  \eta_{\beta \gamma} X_{\alpha \delta} +\eta_{\alpha \delta} X_{\beta \gamma} 
- \eta_{\alpha \gamma} X_{\beta \delta} - \eta_{\beta \delta} X_{\alpha \gamma}. 
\end{equation*} 
In the basis 
\begin{align*}
&M_{\mu\nu}=X_{\mu \nu},\\
&P_\mu = \frac{1}{2} (X_{(d+1) \mu} - X_{d \mu}), \\
&D = X_{45},\\
&K_\mu=2 (X_{(d+1) \mu} + X_{d \mu}),\\
\end{align*}
the so$(2,d)$ algebra reads
\begin{align}
&[D,M_{\mu \nu}] = 0,\label{so2debut}\\
&[D, K_\mu] = K_\mu,\\
&[P_\mu, D] = P_\mu,\\
&[P_\mu, K_\nu] = 2 (\eta_{\mu \nu} D - M_{\mu \nu}),\\
&[K_\mu, M_{\nu \rho}] = \eta_{\mu \nu} K_\rho - \eta_{\mu \rho} K_\nu,\\
&[K_\mu, K_\nu] = 0,\\
&[M_{\mu \nu}, M_{\rho \sigma}] = \eta_{\nu \rho} M_{\mu \sigma} +\eta_{\mu \sigma} M_{\nu \rho} 
- \eta_{\mu \rho} M_{\nu \sigma} - \eta_{\nu \sigma} M_{\mu \rho},\\ 
&[P_\mu, P_\nu] = 0,\\
&[P_\rho, M_{\mu \nu}] = \eta_{\mu \rho} P_\nu - \eta_{\nu \rho} P_\mu\label{so2dfin}. 
\end{align}
We remind from \cite{pconf1} that the $(d-1)(d-2)/2$ generators $X_{ij}$ and the $d-1$ generators $X_{0i}$ of so$(2,d)$ leave $X_{\sss H}$ invariant.  The $d$ more generators which also leaves $X_{\sss H}$ invariant  
are found to be
\begin{equation}\label{yh}
Y_\mu := \frac{1}{2} (1 - H^2) X_{(d+1) \mu} - \frac{1}{2} (1 + H^2) X_{d \mu}.
\end{equation}
A straightforward calculation leads to the following commutations relations:
\begin{eqnarray*}
\left[Y_\mu, Y_\nu\right] &=&  H^2  X_{\mu \nu}, \\
\left[ Y_\rho, X_{\mu \nu} \right] &=& \eta_{\mu \rho} Y_\nu - \eta_{\nu \rho} Y_\mu. 
\end{eqnarray*}

\end{document}